\documentclass[prd,
               twocolumn,
               eqsecnum,
               showpacs,
               letterpaper,
               superscriptaddress,
               altaffilletter,
               nofootinbib, nobibnotes
               ]{revtex4-1}
\usepackage[utf8x]{inputenc}
 \usepackage{graphicx}
  \usepackage{amsmath,amssymb} 
 \usepackage[english]{babel}
  \usepackage{rotating}
  \usepackage{amsbsy}
 \usepackage{textcomp}
 \usepackage{psfrag}
 \usepackage{multirow}
 \usepackage{color} 
 \usepackage{sidecap}
 \usepackage{array}
 \usepackage{subfig}

 \newcommand{\beq }{\begin{equation}}
\newcommand{\eeq}{ \end{equation}}
\newcommand{\beqa }{\begin{eqnarray}}
\newcommand{\eeqa }{\end{eqnarray}}
 \newcommand{\bwt }{\begin{widetext}}
 \newcommand{\ewt }{\end{widetext}}
 \newcommand{\bef}{\begin{figure}[h!]}
\newcommand{\eef}{\end{figure}}

\newcommand{\CQG}{\emph{Class. Quantum Grav. }}
\newcommand{\apjl}{\emph{Astrophys. J. Letters }}
\newcommand{\arxiv}{\emph{Arxiv }}
\newcommand{\Mchirp}{\mathcal{M}}
\newcommand{\dchi}{d$\chi_3$ }
\newcommand{\nonPn}{NonPN }
\newcommand{\powerM}{powerM }

\begin{document}
\title{How serious can the stealth bias be in gravitational wave parameter estimation?}
\author{Salvatore Vitale}
\email[email:  ]{salvatore.vitale@ligo.org}
\affiliation{Massachusetts Institute of Technology, 185 Albany St, 02139 Cambridge USA}
\author{Walter Del Pozzo}
\email[email:  ]{wdp@star.sr.bham.ac.uk}
\affiliation{Nikhef - National Institute for Subatomic Physics, Science Park 105, 1098 XG Amsterdam, The Netherlands}
\affiliation{School of Physics and Astronomy, University of Birmingham, Edgbaston, Birmingham B15 2TT, United Kingdom}

\begin{abstract}

The upcoming direct detection of gravitational waves will open a window to probing the strong-field regime of general relativity (GR). As a consequence, waveforms that include the presence of deviations from GR have been developed (e.g. in the parametrized post-Einsteinian approach). TIGER, a data analysis pipeline which builds Bayesian evidence to support or question the validity of GR, has been written and tested.
In particular, it was shown recently that data from the LIGO and Virgo detectors will allow to detect deviations from GR smaller than can be probed with Solar System tests and pulsar timing measurements or not accessible with conventional tests of GR. However, evidence from several detections is required before a deviation from GR can be confidently claimed. An interesting consequence is that, should GR not be the correct theory of gravity in its strong field regime, using standard GR templates for the matched filter analysis of interferometer data will introduce biases in the gravitational wave measured parameters with potentially disastrous consequences on the astrophysical inferences, such as the coalescence rate or the mass distribution. 
We consider three heuristic possible deviations from GR and show that the biases introduced by assuming GR's validity  manifest in various ways. The mass parameters are usually the most affected, with biases that can be as large as $30$ standard deviations for the symmetric mass ratio, and nearly one percent for the chirp mass, which is usually estimated with sub-percent accuracy.
We conclude that statements about the nature of the observed sources, e.g. if both objects are neutron stars, depend critically on the explicit assumption that GR it the right theory of gravity in the strong field regime.
\end{abstract}
\maketitle

\section{Introduction}

Gravitational waves (GW) generated during the coalescence of compact binary systems (CBC) are the only mean to directly probe the space-time in its genuine dynamical regime. Despite the fact that General Relativity (GR) has so far passed all experimental tests with great success \cite{Will2006}, those tests were performed in situations where the field is weak or stationary and the full non-linear dynamics of GR were not explored. 

On the other hand, in compact objects like neutron stars or black holes coalesce, they approach orbital velocities as high as $50\%$ the speed of light, when very close to merging with the companion. 

The LIGO \cite{ILigo,ALigo} and Virgo \cite{IVirgo,IVirgo2,IVirgo3,AVirgo} ground based gravitational waves observatories are currently undergoing major upgrades, and are scheduled to go back online in 2015 and 2016 respectively~\cite{ObserveScenario}, collecting data with a sensitivity that should allow for a few up to a few tens of CBC detections per year. The exact number will depend on the actual sensitivity reached by the instrument, as well as on the formation rate of compact binary systems, which is still rather uncertain \cite{rates}.
The worldwide network of GW detectors will continue expand during the decade, with LIGO India \cite{Indigo} and Kagra \cite{LCGT} joining operations by 2020. Additional instruments will dramatically improve the sky localization accuracy of GW sources, as well as increase the number of detectable sources \cite{Schutz:2011}.

The prospect of frequent detections calls for the use of a Bayesian framework, which allows the information from each signal to be used to either infer some underlying general property of the observations (see for example \cite{delpozzo12,tayloretal12} for applications in the context of cosmology), or accumulate evidence for a specific model to explain the observations (see \cite{delpozzoetal13} in the context of measuring the neutron star equation of state and \cite{lietal12a} in the context of tests of GR). 

Regarding the strong field deviations from GR, the following questions seem interesting:

\begin{enumerate}
 \item Will LIGO and Virgo be able to confidently recognize a deviation from GR in a detected signal?\label{Item.Tiger}
 \item Should a deviation from GR be visible, wil it be possible to associate the GW signal with a given alternative theory of gravity?\label{Item.OnProgress}
 \item Should a deviation be present but not taken into account in the analysis, how would this affect the estimation of physical (e.g. source masses) and extrinsic (e.g. distance) parameters of the source? \label{Item.This}
\end{enumerate}

The answers will obviously depend on the actual nature and magnitude of the deviation from GR, which may be one of the existing proposed alternative theories (see \cite{cornishetal11} and references therein for a list of alternative theories of gravity) or something unanticipated. 

Some literature exists which answers the point \ref{Item.Tiger} above: the authors and collaborators have shown \cite{lietal12a,lietal12b} how advanced LIGO and Virgo will be sensitive to quite generic (heuristic) deviations from GR. 
They have built a pipeline (TIGER) which works with \emph{any kind} of deviation from GR, thus not requiring the data analyst to know the deviation's form. 
Its efficiency was tested by simulating several kinds of deviations from GR, of comparable magnitude, and it has been found that deviations from GR will eventually be evident by combining evidence (in a Bayesian framework) from several signals.

The exact number of detections required to confidently claim a deviation from GR, depends on what deviation was added, but it usually is $\mathcal{O}(10)$.
Even though this proves that non-GR effects beyond solar system and pulsar tests can be measured, it implies that the biases of the point \ref{Item.This} above will be present not only for unmeasurable deviations (\emph{stealth bias})\footnote{One may wonder whether a deviation so small that it can be hardly measured using model selection can have large effects on the estimation of the GR parameters. The answer is usually yes, as model selection will only work if the extra likelihood gained by taking the non GR parameters into account is higher than the penalty paid for having extra parameters (Occam Razor). Because GW detections are noisy, the parameter estimation code will usually be able to shift the GR parameters, hence the bias, to accommodate the deviation.}, but also for deviations whose measurability requires building up evidence with several signals. 
That implies that, should a deviation from GR be present, it may not be discovered immediately, and GR waveforms might be used for the first few detections, introducing a bias in the parameter estimation process. 

The idea of ``fundamental bias'', i.e. bias in estimated gravitational wave parameters induced by the assumption that GR is  correct, was first introduced in \cite{yunespretorius09}. Subsequently, \cite{cornishetal11} coined the term ``stealth bias'' to describe the class of fundamental biases that cannot be corrected a posteriori, since the data do not provide enough evidence to favor an alternative theory of gravity. 
With the use of analytical approximations, \cite{vallisneriyunes13} follows up by exploring the conditions, expressed in terms of signal-to-noise (SNR) ratio and magnitude of the deviation from GR, in which single events will be affected by stealth bias. They conclude that significant systematic bias might occur, even for deviations that are not yet excluded by observational constraints.

However, \cite{vallisneriyunes13} takes an approach which is valid for loud signals. Because we expect the sources to be distributed uniformly in co-moving volume, the majority of the gravitational wave events will be weak, with SNR close to the threshold necessary for a confident detection. 
For this reason, a full numerical study as close to the real data analysis process as possible is necessary to get more general answers which are valid for signals in a broad range of SNR and parameters.

In this paper, we investigate the transition regime from stealth bias to fundamental bias, considering heuristic deviations from GR that are too small to be confidently detected from any single source observation, but that can be detected after multiple detections. In particular, we focus on the inferences that can be drawn about the class of observed systems from the measurement of the masses. We find that, before enough evidence is accumulated to detect a deviation from GR, the mass measurements can be heavily biased when measured with GR templates. We thus recommend that any astrophysical conclusion drawn by gravitational waves observations should be explicitly conditional on the validity of GR.

The rest of this paper is structured as follows. In section \ref{Sec.Method} we present the experimental set up and the example deviations from GR that were considered; in section \ref{Sec.Results} we present results for each of the deviations; finally, in section \ref{Sec.Conclusions} we draw conclusions and discuss our findings.

\section{Method}\label{Sec.Method}

The bias induced by an unaccounted for deviation from GR in the detected gravitational wave will strongly depend on the exact shape and magnitude of the deviation. Thus, to perform our analysis, we had to choose how to modify the GW signals. While the most natural choice would have been to select some proposed alternative theories of gravity, we have decided not to do so. The reason is twofold: (i) at the moment of writing (with the exception of the investigation of a Massive Gravity theory in \cite{delpozzoetal11,cornishetal11}) no full Bayesian analysis has been performed to check to what extent those theories can be confirmed or ruled out with GW observations; (ii) the class of alternative theories for which usable waveforms are available is very limited, thus limiting the scope of our investigation. 

Particularly in view of (ii), we have picked some of the heuristic GR deviations investigated in \cite{lietal12a,lietal12b}.  
We chose deviations that are detectable, but only when evidence from $\mathcal{O}(10)$ detections is accumulated. Until then, the analysis with standard GR templates may not be unambiguous and, when a detection is made, no final statements may be made about, e.g., the nature of the source (e.g. a binary black hole (BBH) vs a black hole - neutron star (NSBH) or a binary neutron star (BNS)).

Specifically, in addition to standard GR waveforms, we have considered three possible deviations from GR:

\begin{enumerate}
 \item $10\%$ deviation in the 1.5 post-Newtonian (PN) phase coefficient (\cite{lietal12a},~IV~A.1);
 \item an extra term in the phase of the GW, corresponding to a ``1.25'' PN order (\cite{lietal12a},~IV~C);
 \item an extra term in the phase, whose frequency content depends on the total mass of the binary system (\cite{lietal12b},~3.1);
\end{enumerate}

The first step was to generate a catalog of $150$ events, with component masses uniformly distributed in the range $[1.2-2.8] M_\odot$. The position and orientation parameters were uniformly distributed on the unit sphere, while the distances were distributed uniformly in co-moving volume, in the distance range $[50-400]$ Mpc, keeping only events with network SNRs in a realistic range [10-25].

In addition to the standard GR catalog, three catalogs of non-GR signals were generated by assuming the same events as in the GR catalog but adding the deviations described in the aforementioned list. Henceforth we will refer to those modified catalogs as \emph{\dchi} (1.5PN deviation), \emph{\nonPn} (``1.25PN'' deviation) and \emph{\powerM} (deviation with a mass dependent power of the frequency).
All the signals in the four catalogs (three non-GR plus one GR), were thus identical except for the eventual non-GR contribution to the phase.

Each signal in the four catalogs was analyzed using \verb+lalinference_nest+, \cite{S6PE,lalinfPaper}, a Bayesian parameter estimation code based on the Nested Sampling \cite{Skilling} algorithm. The analysis for the three non-GR catalogs were performed using GR templates, thus simulating the situation in which a non-GR deviation is present in the waveform, but it is not accounted for in the analysis.
We also analyzed the GR catalog (using GR templates) to have an idea of the typical uncertainties and (eventual) biases due to poor sampling, noise, etc., in the ``optimal'' case when the template perfectly matches the injected\footnote{We use here the LIGO/Virgo jargon whereby injection means the process of adding a simulated signal into the noise data stream.} signal.

Each of the simulated waveforms was added to zero mean stationary Gaussian noise with a power spectral density corresponding to the design sensitivity of advanced LIGO and Virgo \cite{commissioningdocument}, as coded in the \verb+lalsimulation+ library \cite{LAL}.
The signals were generated using the so-called TaylorF2 approximant (\cite{BuonannoOverlap}), as produced by the \verb+lalsimulation+ package which is part of the LIGO Algorithm Library \cite{LAL}, considering phase contributions up to the 3.5 PN order (O PN in amplitude). No spins were considered in the waveform due to computing limitations. Generic spins would force us to use time domain waveforms (e.g. SpinTaylorT4, \cite{BuonannoOverlap}) which are much slower to calculate, making it impractical for our large scale research program. 

The TaylorF2 approximant is written in the frequency domain, as:

\beq
h(f) = \frac{1}{D} \frac{\mathcal{A}(\theta,\phi,\iota,\psi,\mathcal{M},\eta)}{\sqrt{\dot{F}(\mathcal{M},\eta;f)}} f^{2/3}\,e^{i\Psi(t_c, \phi_c, \mathcal{M},\eta;f)},
\label{Eq.TaylorF2}
\eeq
where $D$ is the luminosity distance to the source, $(\alpha,\delta)$ are right ascension and declination, $(\iota,\psi)$ give the orientation of the orbital plane with respect to the line of sight, $\Mchirp$ is the chirp mass, and $\eta$ is the symmetric mass ratio. They are defined in terms of the component masses $(m_1,m_2)$, as: $\eta = m_1 m_2/(m_1 + m_2)^2$ and $\Mchirp = (m_1 + m_2)\,\eta^{3/5}$. $t_c$ and $\phi_c$ are the time and phase at coalescence, respectively. 
$\dot{F}(\mathcal{M},\eta;f)$ is an expansion in powers of the frequency $f$ with coefficients that depend on mass (and eventual spins) and
\beqa
\label{psiidef}
\Psi^{\mathrm{GR}}(t_c, \phi_c, \mathcal{M},\eta;f)&=& 2\pi f t_c - \phi_c - \pi/4  \\
&+& \sum_{i=0}^7 \left[\psi_i + \psi_{i}^{(l)} \ln f \right]\,f^{(i-5)/3}.\nonumber
\label{phase}
\eeqa

The explicit forms of the coefficients $\psi_i$ and $\psi_{i}^{(l)}$ in $(\mathcal{M},\eta)$ are given, for example, in \cite{mais10}. 

Each of the signals in the non-GR catalogs had an extra term added to the phase. These were:

\begin{enumerate}
 \item \dchi: For the \dchi  deviation, the 1.5 PN phase terms were shifted by 10\%, $\psi_3 \rightarrow \psi_3 (1+0.1)$.
 That is: $\Psi = \Psi^{\mbox{GR}} + 0.1 \frac{3}{128 \eta} (-16\pi) (\pi \Mchirp f)^{-\frac{2}{3}}$
 \item \nonPn: In this case the power of the $(\pi \Mchirp f)$ term is not normally present in the PN series (that would be $[i-5]/3$ with i an integer): $\Psi = \Psi^{\mbox{GR}} - 2.2 \frac{3}{128 \eta} (\pi \Mchirp f)^{-\frac{5}{6}}$. The prefactor $-2.2$ was chosen to make the magnitude of this deviation comparable to the previous one at a reference frequency of 150Hz for a system of $1.5-1.5 M_\odot$. 
 \item \powerM: Finally, for the \powerM catalog, the power of $(\pi \Mchirp f)$ was a function of the total mass of the system $M$: $\Psi = \Psi^{\mbox{GR}} +\frac{3}{128 \eta} (\pi \Mchirp f)^{-2+\frac{M}{3 M_\odot}}$. Here again the prefactor was such that the magnitude of this deviation is comparable to the other two, at the same reference frequency and mass. 
\end{enumerate}

The reader is referred to~\cite{lietal12a,lietal12b} for more details about these deformation, their magnitude and their measurability with Advanced LIGO and Virgo.

\section{Results}\label{Sec.Results}

\subsection{GR injection with GR recovery }

In this section we describe the performance of the parameter estimation (PE) process in the case where both injection and template obey GR. This will serve as a reference for the analysis of non-GR signals. The results will be presented in detail because, to the best of our knowledge, a systematic and statistically large set of events analyzed with the ``full'' PE pipeline (as opposed to Fisher Matrix results, e.g. \cite{Vitale2011}) is not present in the literature. 

It is a common assumption that the chirp mass $\Mchirp$ is well estimated, with sub-percent relative errors (e.g. \cite{Vitale2011,zanolinetal10,arunetal05,arunetal05erratum} with Fisher Matrix, \cite{S6PE} with the MCMC code), while the component masses are estimated with $\mathcal{O}(10)$ percent error, which will make it hard to infer the nature of the source (i.e. whether it was a BNS, NHBS or low mass BBH) \cite{hannametal13}.

Our findings confirm this assumption; the chirp mass is estimated with relative errors (i.e. standard deviation divided by the injected value: $\Gamma_\alpha\equiv \sigma_\alpha/\alpha_{\mbox{true}}$) that are never larger than $\pm0.1\%$ across the whole SNR range, while typical values are $\sim 0.03-0.04\%$ . This is shown in Fig. \ref{Fig.MchirpBPGRGR}: The boxes are logarithmically spaced to take into account the fact that there were more events at low SNR than at high SNR, so that at least 10 events (unless otherwise indicated) are contained in each box. 
\bef
 \includegraphics[width=\columnwidth]{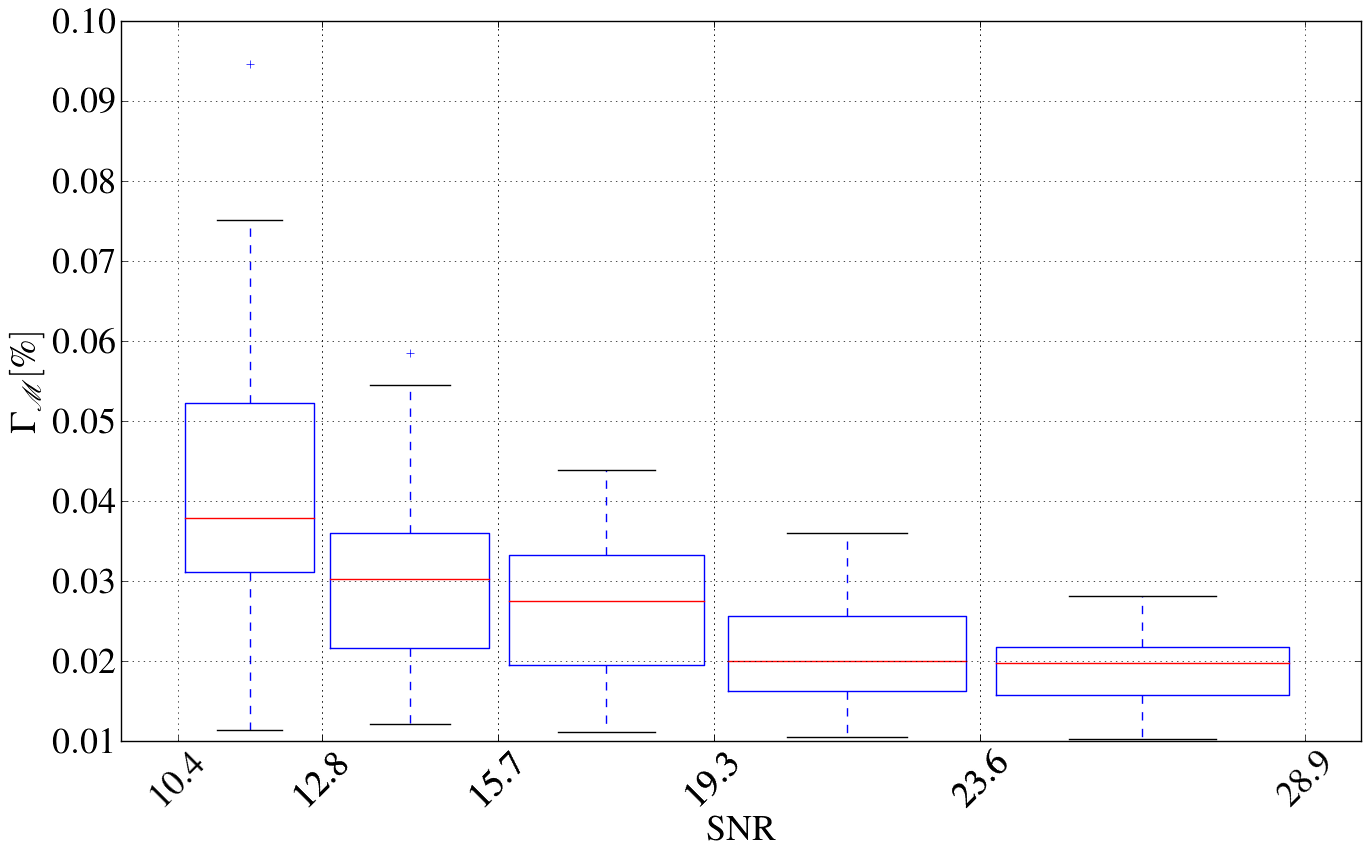}
\caption{Chirp mass percentage relative error, $\Gamma_\Mchirp$, for the GR catalog set of events as a function of the optimal SNR. The boxes indicate the lower to upper quartile values of the data, with a line at the median. The whiskers show 1.5 times the interquantile range, while the symbols are the remaining data points. In every SNR bin under consideration the relative error for $\Mchirp$  is always smaller than $0.1\%$, and usually $\lesssim 0.03\%$ for medium-high SNR signals}\label{Fig.MchirpBPGRGR}
\eef

As for the symmetric mass ratio $\eta = m_1 m_2/(m_1 + m_2)^2$, we find that the error can be as large as $\sim 5\%$, but is generally $\lesssim 2\%$ for medium-low SNR events, and $\lesssim 1\%$ for louder events, as shown in Fig. \ref{Fig.etaBPGRGR}.

\bef
 \includegraphics[width=\columnwidth]{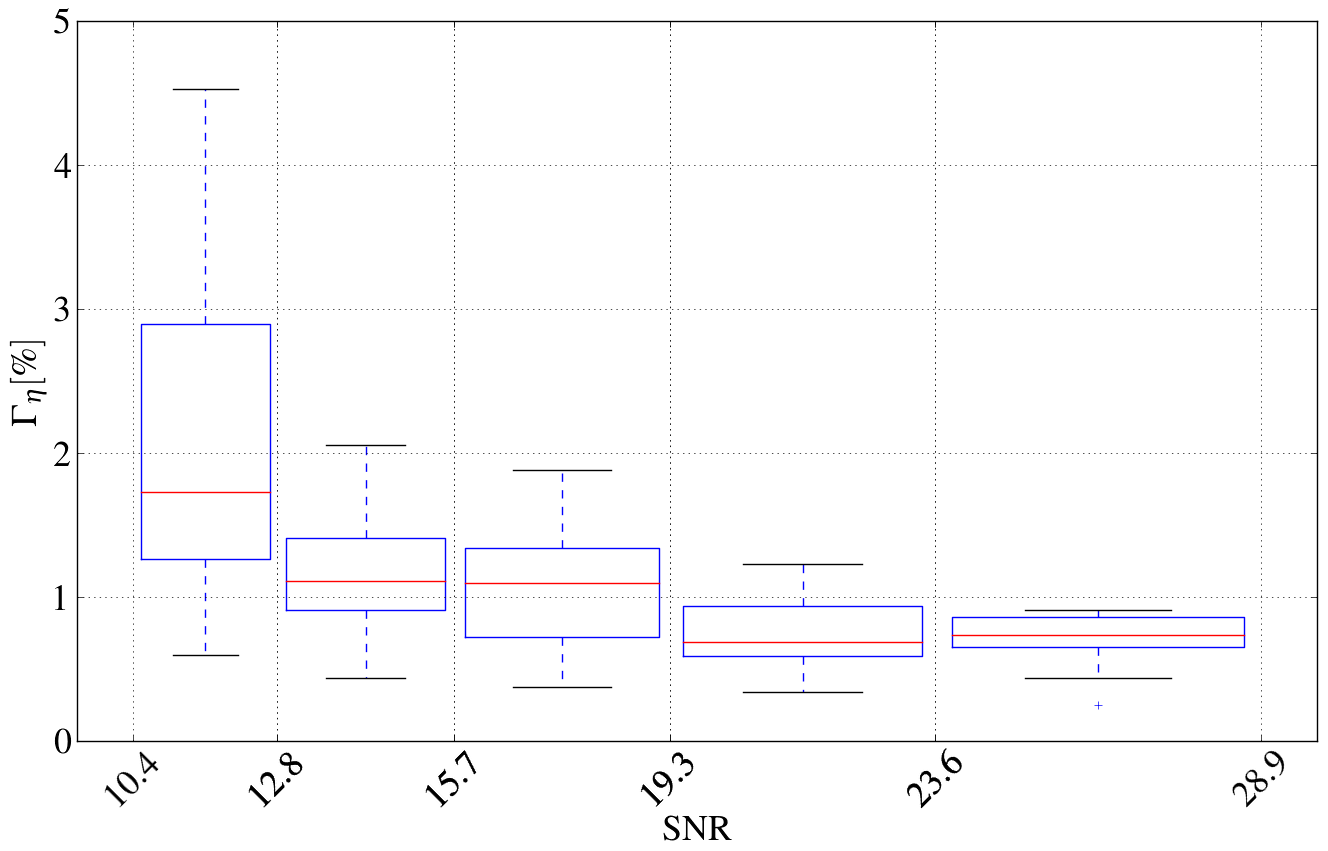}
\caption{Symmetric mass ratio percentage relative error, $\Gamma_\eta$, for the GR catalog set of events as a function of the optimal SNR. The bias can be as large as $\sim 5\%$, but, in general, are $\sim 2\%$ events with medium SNR ($\sim 15$), and $\lesssim 1\%$ for louder event.}\label{Fig.etaBPGRGR}
\eef

In particular, we find that for $50\%$ ($90\%$) of the signals the relative error for the symmetric mass ratio is smaller than $1\%$ ($2\%$).

We would like to draw some conclusions about how often the PE code will make the correct inference on the nature of the source, i.e. how precisely the component masses can be estimated.
Because the maximum mass of a neutron star is a function of the yet unknown equation of state, e.g. \cite{lattimer12}, we recognize that, for the time being, there can \emph{not} be an uncontroversial choice for this upper bound.
We thus choose the reasonable value of $2.0 M_\odot$ \cite{AntoniadisAl}. Henceforth, we will label as ``neutron star'' (``black hole'') an object lighter (heavier) than $2.0M_\odot$.

Having made clear our choice and its limitations, we can now calculate how often a system that was injected into a mass bin, e.g. BNS, is correctly assigned to the corresponding mass bin at a two sigma confidence level.
We must also consider the possibility that we will not be able to make a decision (i.e. that the error bars on the component masses are such that we cannot decide where to put the signal) or that a wrong inference will be made, assigning the injection to a different mass bin than where it was injected. This point will become more important when non-GR injections will be performed.

Our findings are reported in Table. \ref{Tab.GR_GR}: nearly half of the signals are assigned to the correct mass bin, at 95\% confidence level.
Even more important, we notice that signals are either assigned to the correct mass bin or not assigned at all, and that none of the signals are assigned to the wrong mass bin. We will see that the situation is very different whit non-GR injections. 

\begin{table}[h]
\begin{tabular}{|c|c|c|c|c|c|}
\hline
&&\multicolumn{3}{c|}{Rec. as}&\multicolumn{1}{c|}{Unassigned}\\
&& BNS & NSBH & BBH & at $2\sigma$\\
 \hline
\multirow{3}{3mm}{\begin{sideways}\parbox{15mm}{Inj. as}\end{sideways}}& BNS & 45\%& 0 & 0&55\%\\ [1.1ex] \cline{2-6}
&NSBH & 0 & 44\% & 0&56\% \\ [1.1ex]\cline{2-6}
&BBH   & 0 & 0 & 48\%&52\% \\ [1.1ex]\cline{2-6}
\hline
\end{tabular}
\caption{Fractions of signals recovered as a BNS, NSBH or BBH at two sigma confidence level (see the text for details). Nearly half of the times the code is able to infer correctly what had been injected. \label{Tab.GR_GR}}
\end{table} 

We also have checked how the efficiency in assigning injections to the correct mass bin depends on the injected chirp mass. As expected, the efficiency is quite high ($\sim60\%$ for low mass BNS and $\sim 90\%$ for high mass BBH) for systems which are either very light or very heavy, and decreases significantly for systems with chirp mass in the range $\sim 1.4-2.0$ where the error bars on the component masses can easily cross the BNS-NSBH or NSBH-BBH boundary, and no decision can be made.
A summary of the GR analysis is reported in Table~\ref{Tab.GR_stat}. The first five rows reports statistics for 
for $\Mchirp,\eta,m_{1,2},D$ and $\iota$.
We quote the median and $90^{\mbox{th}}$ percentile of the relative errors $\Gamma$ and the absolute value of the \emph{effect size} $\Sigma$, defined as:
\beq\label{Eq.Esize}
\Sigma_\alpha\equiv\frac{\bar{\alpha} - \alpha_{\mbox{true}}}{\sigma_{\alpha}};
\eeq

where $\bar{\alpha}$ and $\sigma_\alpha$ are the median and standard deviation of the posterior distribution of $\alpha$, and $\alpha_{\mbox{true}}$ is the injected ``true'' value. The effect size represents the offset in units of standard deviation.

For all parameter, 50\% (90\%) of signals have a median which is found within $\sim 0.6$ ($\sim 1.5$) standard deviations. These values are consistent with the expectations for random Gaussian variables, thus showing the robustness of the parameter estimation algorithm. 
The bottom row of Table~\ref{Tab.GR_stat} focuses on sky localization performances: the first two columns report the median and  $90^{\mbox{th}}$  percentile for the 90\% confidence level sky error area (in square degrees), the numbers in brackets correspond to a selection of events having SNR above 8 in \emph{all} three instruments. Finally, the last two columns report the angle (in degrees) between the injected sky position and the maximum likelihood point; 90\% of the signals that had been detected in the three interferometers were found at less than 2 degrees from the true position, and with a sky area  smaller than 33 $\mathrm{deg}^2$.

\begin{table}[h]
\begin{tabular}{|c|c|c|c|c|}
\hline
&\multicolumn{2}{c|}{$\Gamma$}&\multicolumn{2}{c|}{$|\Sigma|$}\\
& 50\% & 90\% & 50\% &90\%\\
\hline
$\Mchirp$ & 0.03\%& 0.05\% & 0.6&1.4\\ [1.1ex] \cline{1-5}
$\eta$ & 1.1\% & 2.2\% & 0.5&1.5 \\ [1.1ex]\cline{1-5}
$m_{1,2}$ &$4.5$\% & $6.8$\% &$0.7$&$1.4$ \\ [1.1ex]\cline{1-5}
$D$ & 20.7\% & 31.3\% & 0.6&1.5 \\ [1.1ex]\cline{1-5}
$\iota$ & 32.5\% & 104.7\% & 0.6&1.6 \\ [1.1ex]\cline{1-5}
\hline\hline
&\multicolumn{2}{c|}{Sky Error [deg$^2$]}&\multicolumn{2}{c|}{Sky offset [deg]}\\
& 50\% & 90\% & 50\% &90\%\\
$\delta\Omega_{90}$ & 15.3 (4.5) & 95.3 (33.3) & 1.7 (0.8) &25 (2.2) \\ [1.1ex]\cline{1-5}
\hline
\end{tabular}
\caption{Summary of errors for the GR analysis. The first two columns, $\Gamma$, quote the $50^{\mbox{th}}$ and $90^{\mbox{th}}$ percentile for the relative errors. The last two columns, $|\Sigma|$, quote the $50^{\mbox{th}}$ and $90^{\mbox{th}}$ percentile for the effect size, eq.\ref{Eq.Esize}. The numbers for each component mass are similar, thus we quote their means in the $m_{1,2}$ row. For the sky localization accuracy (last row) the last two columns report the percentiles on the angle offset between the injected and median recovered sky position. The numbers in brackets refer to events which have an SNR above 8 in all detectors. \label{Tab.GR_stat}}
\end{table} 

The following subsections will be devoted to analyses of the non-GR catalogs. As a general statement, because all the variations from GR we considered affect the phase of the waveform, we would expect that intrinsic parameters, i.e. the mass parameters, are the most affected by the unaccounted deviation. We will see that this is generally the case.

\subsection{\dchi injection with GR recovery }

In this section we report the analysis of the \dchi catalog, i.e. the signals in which a 10\% deviation in the 1.5PN phase terms is present. Here and in what follows we will use two figures of merit for the stealth bias: the relative offset with respect to the injected value:
    
$$
\Delta_\alpha \equiv \frac{\bar{\alpha} - \alpha_{\mbox{true}}}{\alpha_{\mbox{true}}}
$$
   
and the effect size, $\Sigma$, defined in eq. \ref{Eq.Esize}

We found that the chirp mass estimation, Fig. \ref{Fig.MchirpBPdchiGR}, is only mildly affected, with relative offsets that, even if larger (up to $\sim 10$ times) than the typical uncertainties for this parameter in the GR case, are still well below the percent level. As a consequence, the selection of BNS events for a test of GR based on the measured chirp mass \cite{AgathosAl}, will not be affected by stealth bias.

\bef
 \includegraphics[width=\columnwidth]{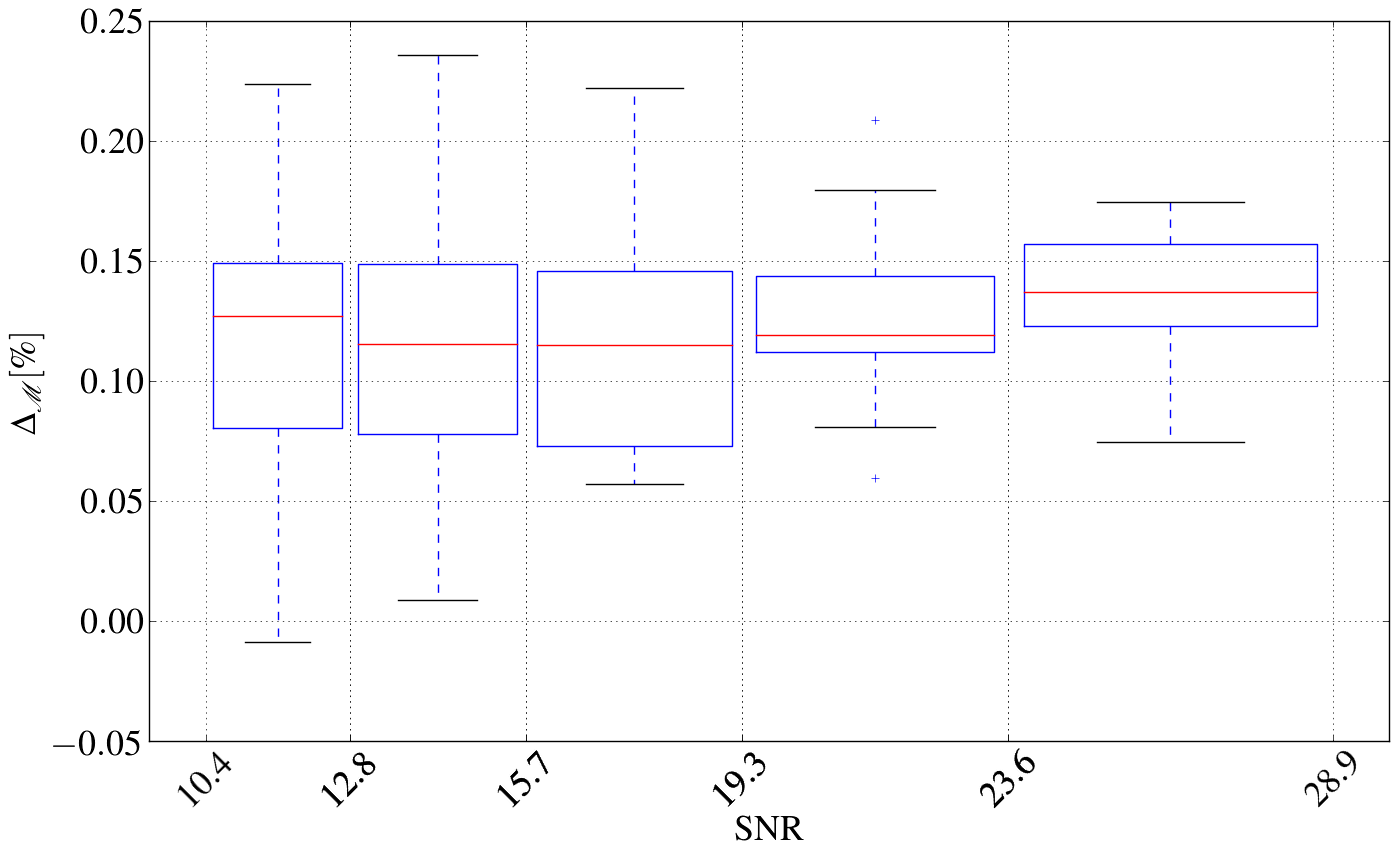}
\caption{$\Delta_\Mchirp$ for the \dchi catalog set of events as a function of the optimal SNR. Line, box, whiskers and symbols are the same as in  Fig.~\ref{Fig.etaBPGRGR}.  In each SNR bin, the median $\Mchirp$ and the injected one differ by $\sim 0.1$\%. While the offset is a factor of three larger than the typical errors for the GR catalog, it is still a fraction of percent.}\label{Fig.MchirpBPdchiGR}
\eef

On the other hand, the mass ratio is heavily affected by the presence of the \dchi deviation. This finding is not totally unexpected, cfr. Fig. 3 in \cite{lietal12a}.
The relative offset is $~ -15 \%$  for all the signals, while the measurement becomes more precise for loud signals.  $\Sigma_\eta$ thus gets larger and larger, with the loudest events being measured $\sim 20 $ standard deviations away from the injected values, Fig.~\ref{Fig.etaBPdchiGR}.

\bef
 \includegraphics[width=\columnwidth]{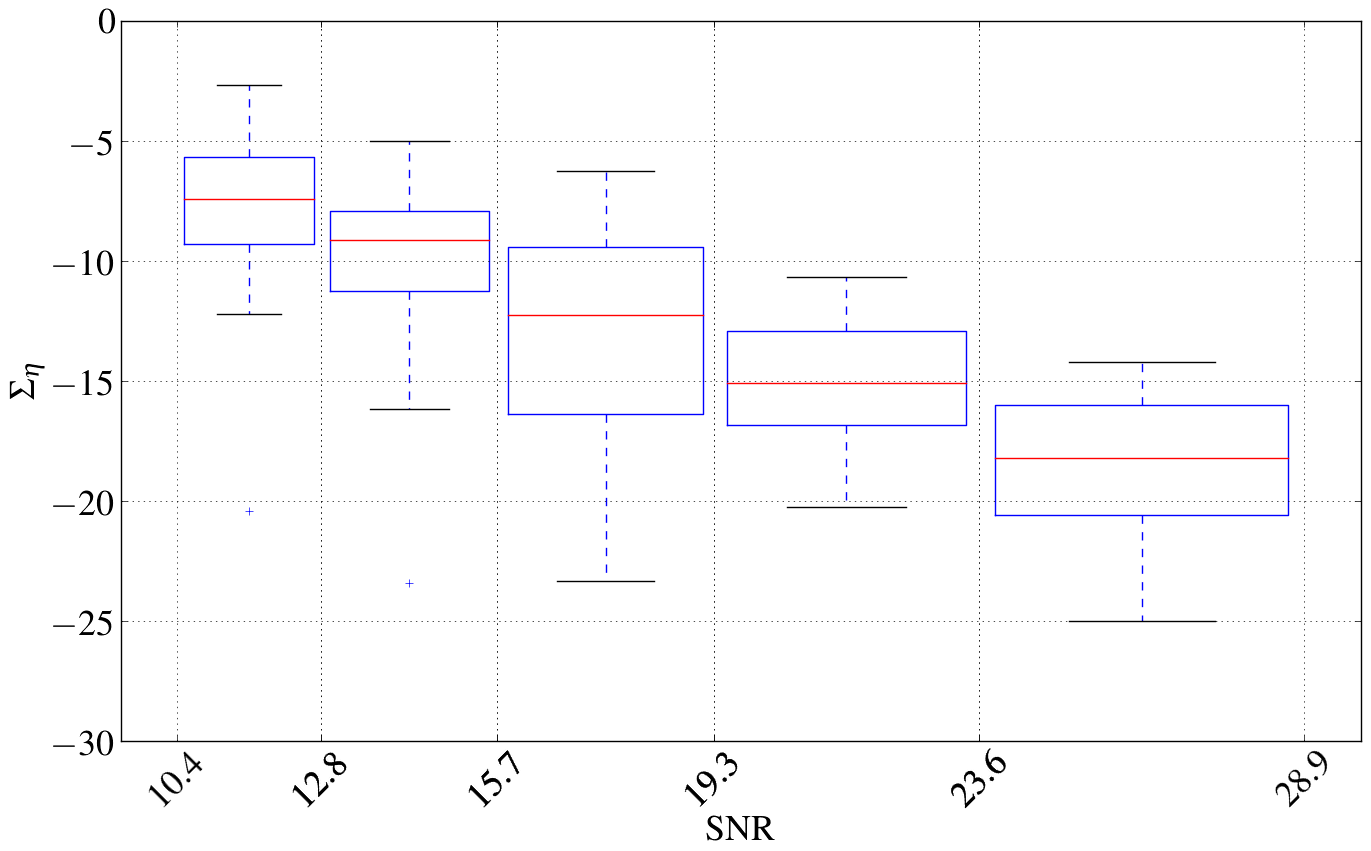}
\caption{$\Sigma_\eta$ for the \dchi catalog set of events as a function of the optimal SNR. Line, box, whiskers and symbols are the same as in  Fig.~\ref{Fig.etaBPGRGR}. The median recovered value is more than 5 standard deviations away from the injected value for low SNR events, and gets further and further as the SNR increases, with the loudest events having $\Sigma_\eta \sim -20$. This is not due to the bias becoming larger for louder signals but to the standard deviation becoming smaller.}\label{Fig.etaBPdchiGR}
\eef

Since the bias is always negative, the parameter estimation algorithm systematically \emph{underestimates} the value of the mass ratio. 
In other words, each system is seen as having component masses that are more different than reality. This is also shown in Table \ref{Tab.dchi_GR} in which the overwhelming majority of the signals are identified as NSBH systems.
Note the difference compared to the GR catalog, Table~\ref{Tab.GR_GR}. In the GR case systems were either not assigned to any mass bin or correctly identified. In the case in which a \dchi departure from GR is present, systems are misidentified 95\% of the times for the BNS case and 98\% of the times for BBH. The NSBH bin is now 100\% (it was 44\% for GR), since all the signals for which a decision could not be made in the GR catalog are being pushed toward very low mass ratios. 
Not a single BNS or BBH gets assigned to the correct mass bin.

\begin{table}[h]
\begin{tabular}{|c|c|c|c|c|c|}
\hline
&&\multicolumn{3}{c|}{Rec. as}&\multicolumn{1}{c|}{Unassigned}\\
&& BNS & NSBH & BBH & at $2\sigma$\\
 \hline
\multirow{3}{3mm}{\begin{sideways}\parbox{15mm}{Inj. as}\end{sideways}}& BNS & 0& 95\% & 0&5\%\\ [1.1ex] \cline{2-6}
&NSBH & 0 & 100\% & 0&0 \\ [1.1ex]\cline{2-6}
&BBH   & 0 & 98\% & 0&2\% \\ [1.1ex]\cline{2-6}
\hline
\end{tabular}
\caption{Fraction of signals recovered as a BNS, NSBH or BBH at two sigma confidence level for \dchi injections. }\label{Tab.dchi_GR}
\end{table} 

The seriousness of the bias is evident from Fig. \ref{Fig.etaComparison} in which we compare the distribution of the injected mass ratios (in red) the posterior medians for the GR (in blue) and \dchi catalog (in green). As expected, in the GR catalog the median estimated values match closely the injected ones. On the contrary, in the \dchi catalog the distribution of the posterior medians is shifted towards smaller values of $\eta$, and has barely any overlap with the distribution of the injected values. The distribution of the recovered symmetric mass ratios for \dchi peaks at $\eta \sim 0.21$ , which means that the most massive star is seen as twice as massive as than the lighter object.

\bef
 \includegraphics[width=\columnwidth]{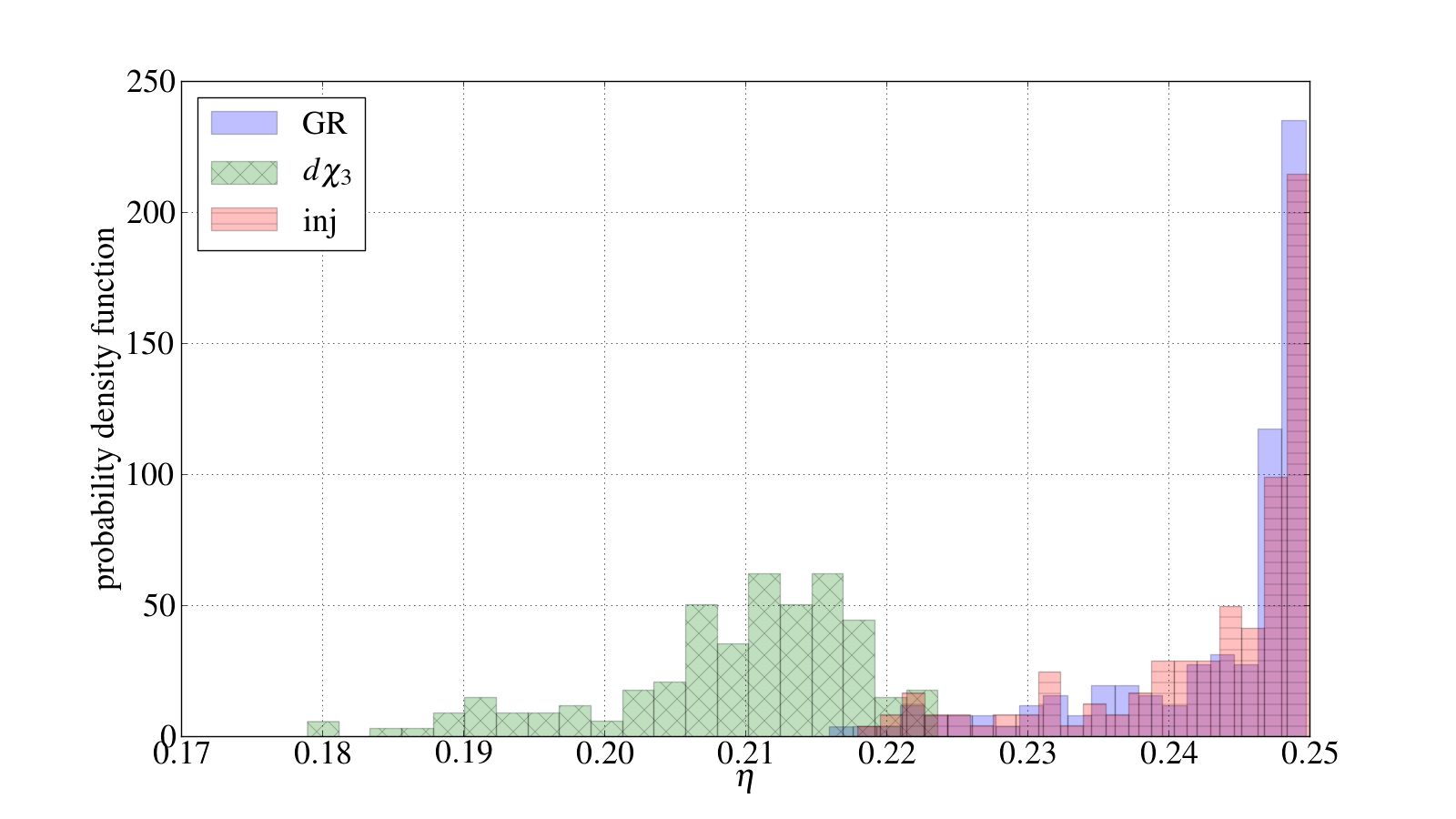}
\caption{Distribution for the injected values of $\eta$ (in red) compared with the distributions of posterior medians as measured in the GR catalog (in blue) and the \dchi catalog (in green). The \dchi distribution is clearly offset from both the injections and the GR.}\label{Fig.etaComparison}
\eef

We further verified whether the bias on the symmetric mass ratio depends on the injected values of the masses of the system, and we found that it does not. Indeed, the non-GR phase terms at a given frequency do not vary significantly over the mass parameters range we are probing here. On the other hand, the \powerM deviation shows a much stronger dependence on the masses, with the magnitude of the shift inversely proportional to the masses. We will indeed see in Sec.~\ref{SubS.PowerM} that the bias is larger for BNS. 

Finally, we remark that, as expected with phase-only deviations, none of the extrinsic parameters shows significant biases with respect to the GR catalog.

\subsection{\nonPn injection with GR recovery }

In this subsection we investigate the \nonPn catalog, for which the deviation from GR predictions can be, with an abuse of notation, dubbed  ``1.25PN''.
For this catalog the relative offset introduced in the chirp mass estimation is moderate, even though generally larger than for the \dchi catalog, being usually around $0.25-0.3$\% for all the event.

We find that the offset in $\eta$ is comparable, in both magnitude and sign, to what was seen for the \dchi runs, i.e. the posterior medians are systematically \emph{underestimated}. This is shown in Fig.~\ref{Fig.etaBPOne25GR}.

\bef
 \includegraphics[width=\columnwidth]{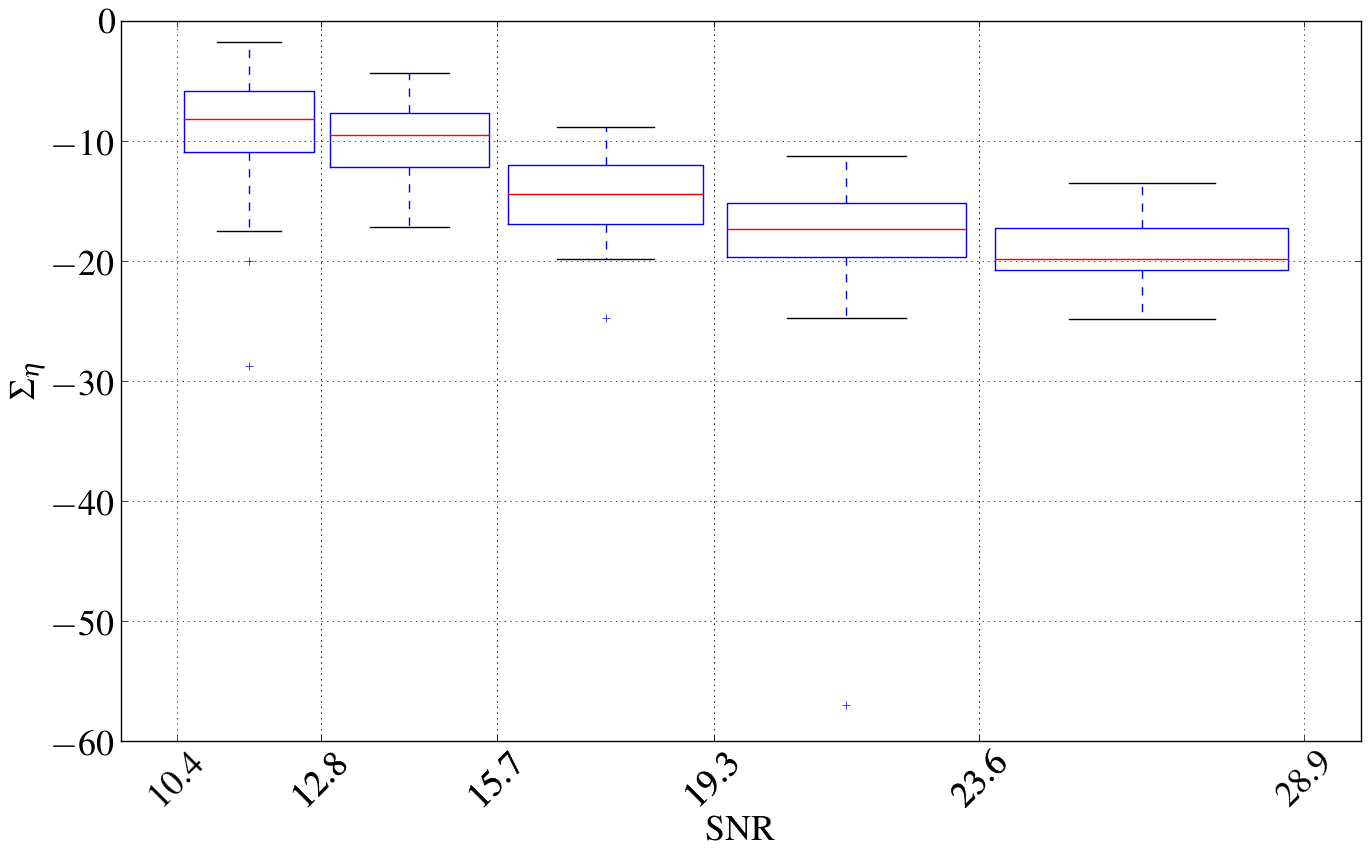}
\caption{$\Sigma_\eta$ for the \nonPn set of events as a function of the optimal SNR. Line, box, whiskers and symbols are the same as in  Fig.~\ref{Fig.etaBPGRGR}. The median recovered value is more than 5 standard deviations away from the injected value for low SNR events, and gets further and further as the SNR increases, with the loudest events having $\Sigma_\eta \sim -20$ }\label{Fig.etaBPOne25GR}
\eef

It is thus not surprising that our findings resemble the $d\chi_3$ catalog. Nearly the totality of events are seen as NSBH, Table~\ref{Tab.nonPN_GR}. 

\begin{table}[h]
\begin{tabular}{|c|c|c|c|c|c|}
\hline
&&\multicolumn{3}{c|}{Rec. as}&\multicolumn{1}{c|}{Unassigned}\\
&& BNS & NSBH & BBH & at $2\sigma$\\
 \hline
\multirow{3}{3mm}{\begin{sideways}\parbox{15mm}{Inj. as}\end{sideways}}& BNS & 0& 100\% & 0&0\\ [1.1ex] \cline{2-6}
&NSBH & 0 & 100\% & 0&0 \\ [1.1ex]\cline{2-6}
&BBH   & 0 & 98\% & 0&2\% \\ [1.1ex]\cline{2-6}
\hline
\end{tabular}
\caption{Fraction of signals recovered as a BNS, NSBH or BBH at two sigma confidence level for \nonPn signals.}\label{Tab.nonPN_GR}
\end{table} 

\subsection{\powerM injection with GR recovery}\label{SubS.PowerM}

In this subsection we describe our findings for the \powerM catalog. As described in Sec.~\ref{Sec.Method}, these signals are characterized by the presence of an extra term in the GW phase whose power of the frequency is a function of the total mass of the system. This is a rather different, and richer, situation than \dchi or \nonPn, where all signals had non-GR shifts with the same frequency dependence, and we may thus expect the resulting bias to manifest itself differently.

We find that the bias in the chirp mass, Fig. \ref{Fig.MchirpBPMGR}, is usually smaller than in the other catalogs, even though several outliers are present for which the offset is several tenths of percent. We also notice that, unlike the other two deviations, the sign of the bias is not the same for all events, but tends to be negative for low-mass events and slightly positive for the most massive sources. Such an effect should not come as a total surprise, given that, by its very nature, this deviation strongly depends on the mass of the system. 

\bef
 \includegraphics[width=\columnwidth]{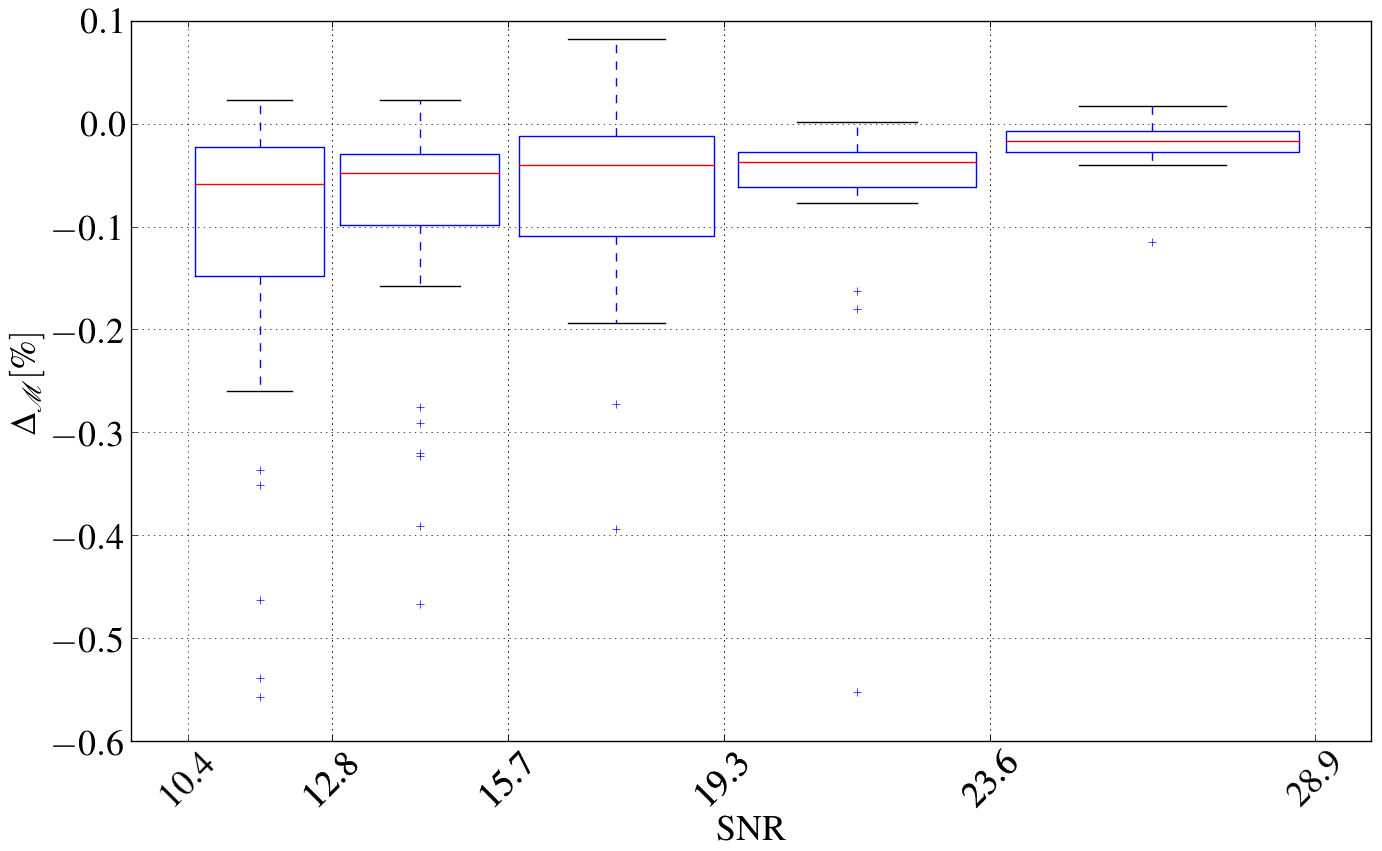}
\caption{$\Delta_\Mchirp$ for the \powerM catalog set of events as a function of the optimal SNR. Line, box, whiskers and symbols are the same as in  Fig.~\ref{Fig.etaBPGRGR}.}\label{Fig.MchirpBPMGR}
\eef

As for $\eta$, Fig.~\ref{Fig.etaBPMGR}, we observe that the sign of the bias is now positive. Component masses will thus be seen as more equal than they actually are.

\bef
 \includegraphics[width=\columnwidth]{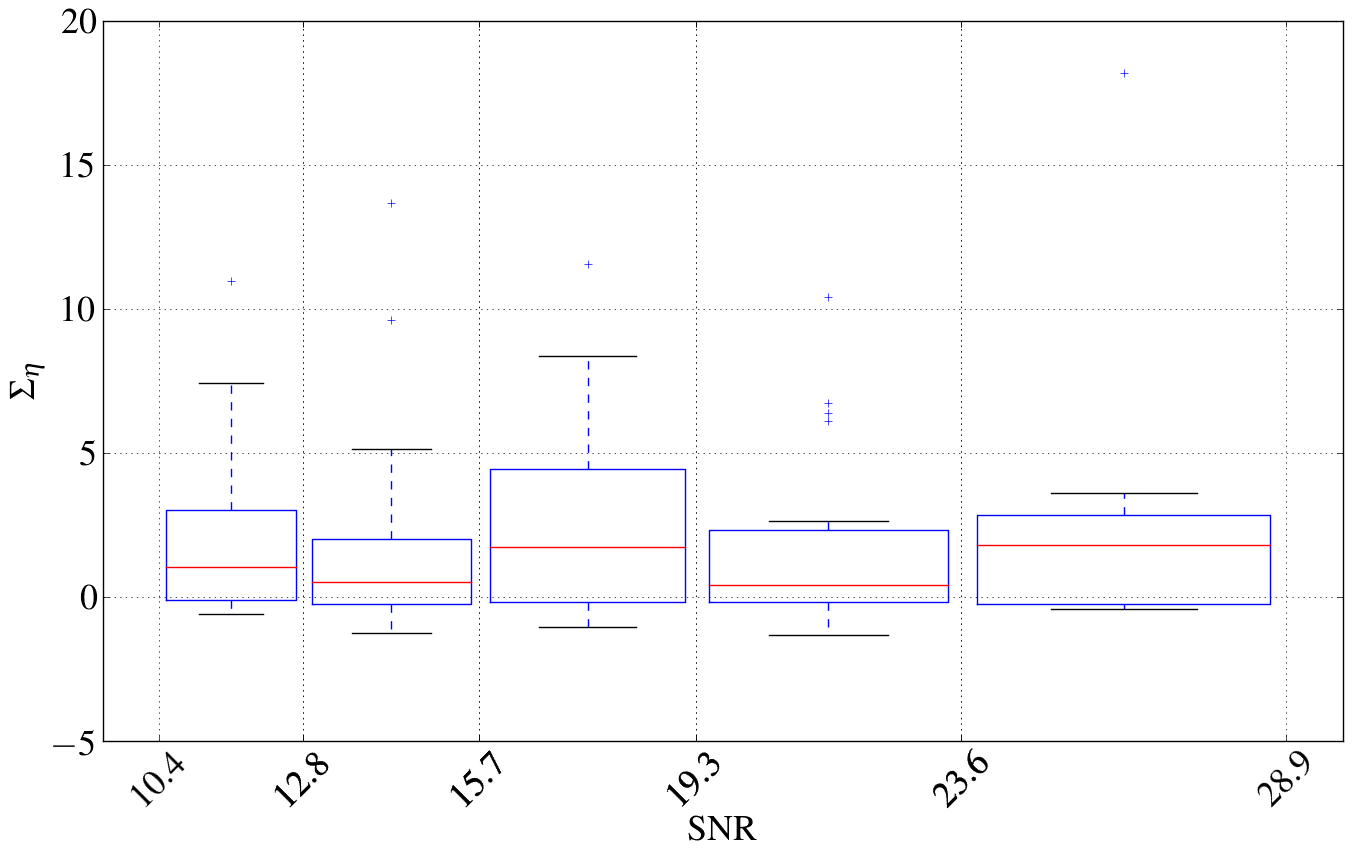}
\caption{Effect size for $\eta$ in the \powerM catalog set of events as a function of the optimal SNR. Line, box, whiskers and symbols are the same as in  Fig.~\ref{Fig.etaBPGRGR}. The effect is generally smaller than what seen in the \dchi or \nonPn catalogs, even if several outliers exist, which are found at $\gtrsim 5$ sigmas from the injected value. We also notice the presence of both positive and negative biases, the sign depending on the injected chirp mass (see text).}\label{Fig.etaBPMGR}
\eef

That is shown in Table~\ref{Tab.powerM_GR}: 64\% of BNS injections are now being recovered as BNS. That is more than in the GR injections case, where the number was 45\%. The explanation is that some of the events for which a decision could not be made for GR injections have been pushed up to higher $\eta$, to the BNS cell.
The same line of thought applies to the NSBH injections: 44\% of them were correctly recognized in the GR catalog while only 29\% are still seen as NSBH, and 9\% are mislabeled as BNS.
Finally, we do not see much difference for the BBH injections, which is due to the fact that the magnitude of the \powerM non-GR shift gets very small for chirp masses $\gtrsim 2 M_\odot$ (see Fig.~\ref{Fig.MchirpEtaScatMGR}, top panel).

\begin{table}[h]
\begin{tabular}{|c|c|c|c|c|c|}
\hline
&&\multicolumn{3}{c|}{Rec. as}&\multicolumn{1}{c|}{Unassigned}\\
&& BNS & NSBH & BBH & at $2\sigma$\\
 \hline
\multirow{3}{3mm}{\begin{sideways}\parbox{15mm}{Inj. as}\end{sideways}}& BNS & 64\%& 0 & 0&36\%\\ [1.1ex] \cline{2-6}
&NSBH & 9\% & 29\% & 0&62\% \\ [1.1ex]\cline{2-6}
&BBH   & 0 & 0& 50\%&50\% \\ [1.1ex]\cline{2-6}
\hline
\end{tabular}
\caption{Fraction of signals recovered as a BNS, NSBH or BBH at two sigma confidence level for \powerM injections. }\label{Tab.powerM_GR}
\end{table} 

We would thus expect, just from back of the envelop calculations, low-mass events to be more heavily biased than higher mass events; that is indeed what we have found. 
The top panel in Fig. \ref{Fig.MchirpEtaScatMGR} shows the bias for the chirp mass (colorbar) for the various events, labeled by their injected $\Mchirp$ and $\eta$. The circles are proportional to the loudness of the event. It is clear how the bias for the chirp mass strongly depends on the injected $\Mchirp$, and is more important for \emph{low} chirp mass systems. 

The equivalent plot for the mass ratio $\eta$, Fig.~\ref{Fig.MchirpEtaScatMGR} bottom panel, shows dependence on both the injected $\Mchirp$ and $\eta$, with large-$\eta$ events getting a smaller bias on $\eta$. 
That is easily understandable. Because the effect of the \powerM deviations from GR is to \emph{increase} the recovered value of $\eta$, events which were already very close to the upper bound ($\eta=0.25$) will be less biased for the simple reason that their posterior distribution cannot move any higher. For those events the effects of the non-GR shift will mostly be to narrow down the $\eta$ posterior distribution.
\bef
 \includegraphics[width=\columnwidth]{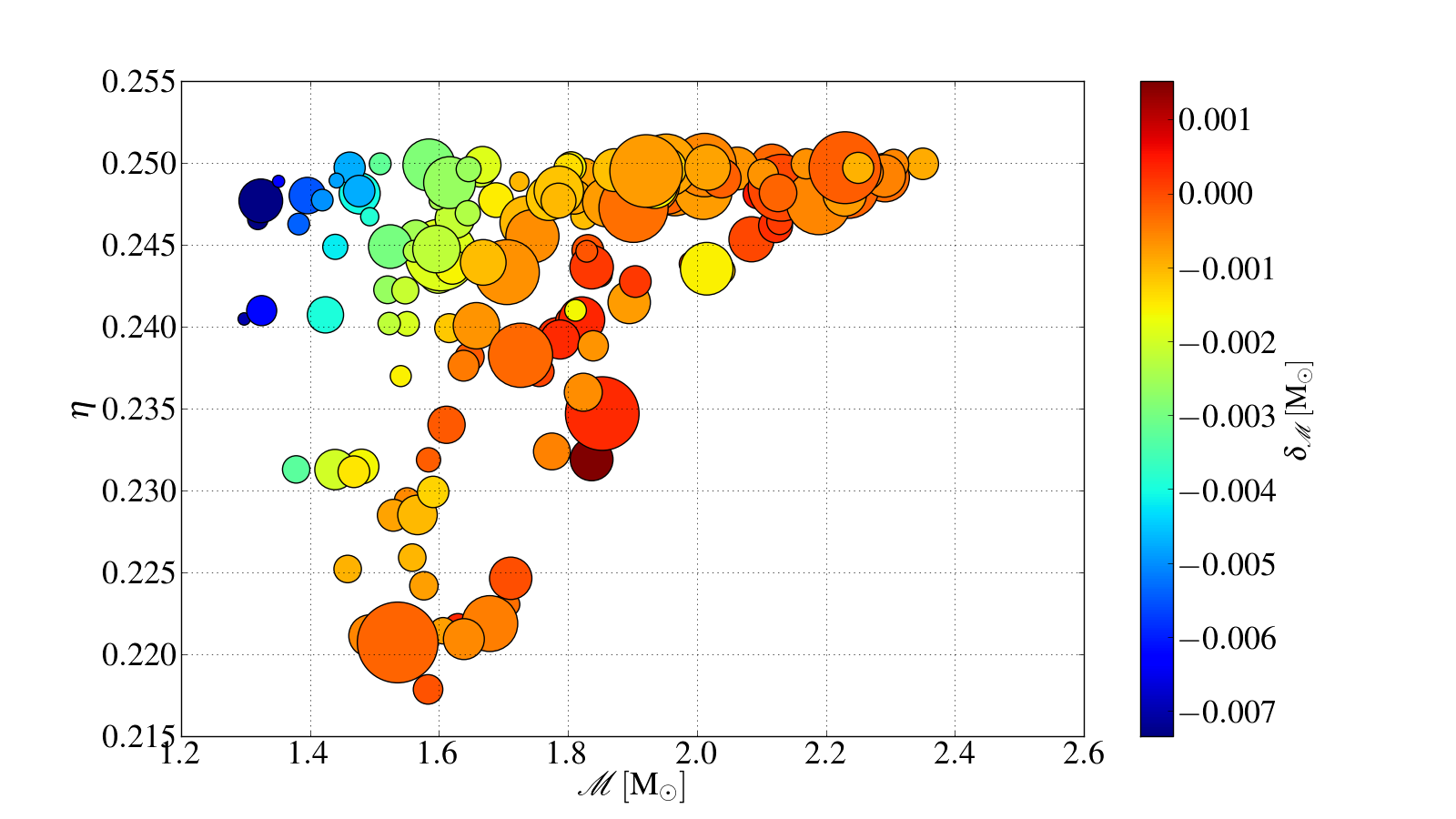} \\
 \includegraphics[width=\columnwidth]{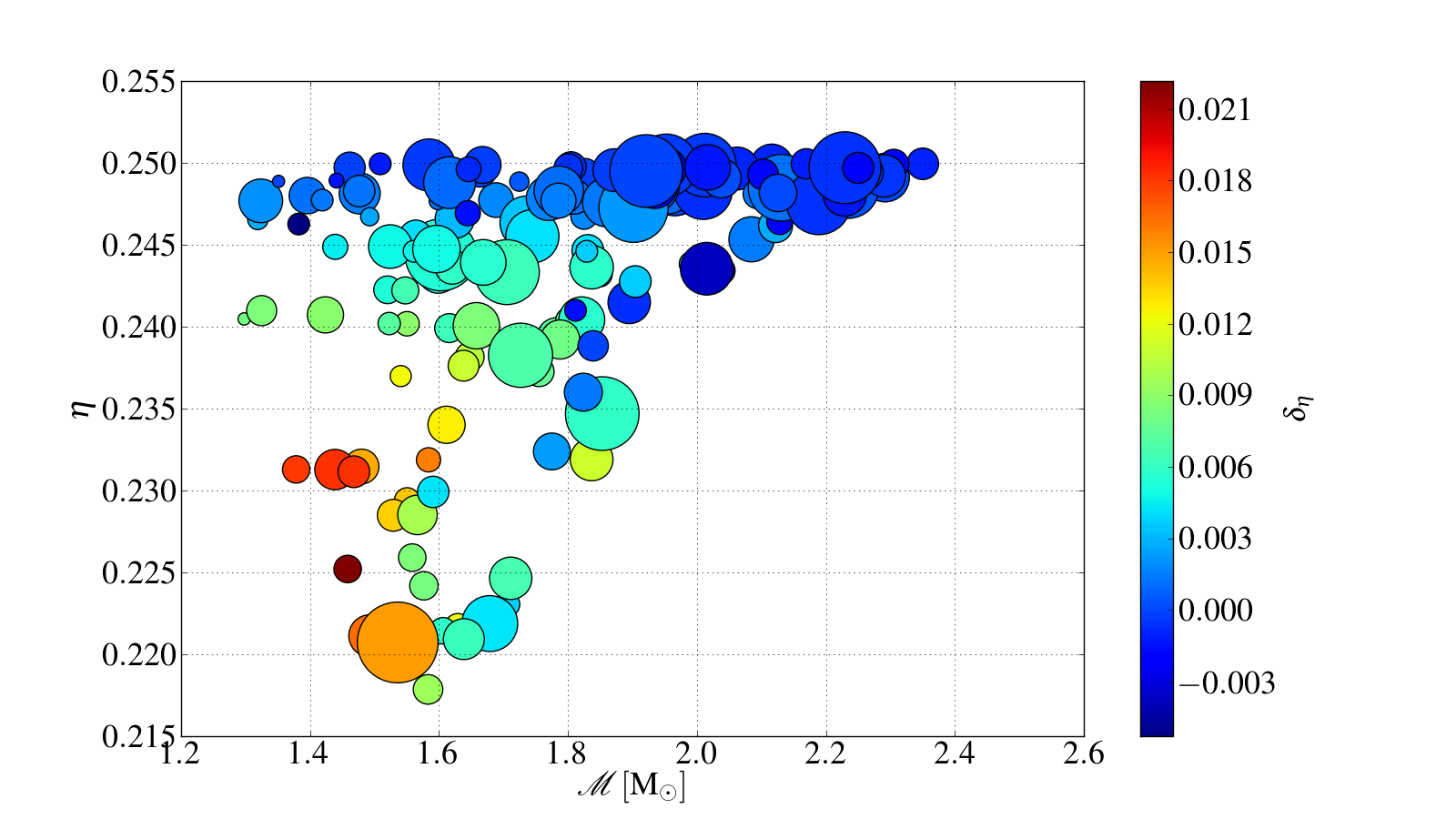}
\caption{Top: bias for the estimation of $\Mchirp$ as function of the injected values of $\Mchirp$ and $\eta$. 
Bottom: bias for the estimation of $\eta$ as function of the injected values of $\Mchirp$ and $\eta$. The bias becomes larger for smaller injected values of $\Mchirp$.}\label{Fig.MchirpEtaScatMGR}
\eef

\section{Conclusions}\label{Sec.Conclusions}

In this work we have shown how parameter estimation for gravitational wave signals can be strongly affected by deviations from general relativity, should they not be taken into account in the waveform used as templates.
For the purpose of illustration, we have considered three heuristic deviations from GR, whose non-GR nature would \emph{eventually} be recognizable using evidence from several, $\mathcal{O}(10)$, detections.

We created a catalog of $150$ signals, that were analyzed a total of 4 times: injecting the ``correct'' GR signal, or injecting one of the 3 deviations from GR.

We have seen that the effect of those deviations on the mass ratio can be very different. It can lead to a heavy systematic underestimation of $\eta$ (\dchi and \nonPn catalog), where the mass ratio is biased toward 2:1, i.e. $\eta\sim 0.22$, or larger, or to an overestimation of $\eta$, \powerM catalog, where the distribution of recovered $\eta$'s is pushed toward the upper boundary at 0.25 and the systems seen as equal-mass.

For the sake of argument, and without claim of astrophysical validity, we have labeled as neutron star (black hole) compact objects lighter (heavier) than $2M_\odot$. We have shown that when GR templates are used for injection and for the analysis, our current parameter estimation algorithms are able to recognize the nature of the injected systems, at the $2\sigma$ confidence level, $\sim$ 50\% of the time.  Moreover, none of the GR signals were assigned to the wrong source class. 

The situation was reversed when injections were allowed to depart from GR and analyzed using GR templates. For the \dchi (\nonPn) catalog, for example, 95\% (100\%) of the injected BNS were mistaken for NSBH. For the \powerM catalog, only $29\%$ of the injected NSBH was recognized as such, while $9\%$ of them were mistaken for BNS.
Even though the numerical details of our findings would change if different mass thresholds were to be chosen, our conclusions can be summarized in two points: 
\begin{enumerate}
\item[i)] if the templates used are a good representation of the detected signals, $\sim 50\%$ of the times we can infer the nature of the detected signal at a $2-\sigma$ confidence level;
\item[ii)] if the templates do not match the signal waveforms well, the measured component masses of a system are an unreliable (and potentially disastrous) indicator of the class of the system.
\end{enumerate}
Therefore, any future inference that will be drawn from an in-depth analysis of GW signals with state-of-the-art parameter estimation algorithms, \emph{is critically and explicitly dependent on the underlying theory of gravity assumed}.  

The study herein reported focused on the bias introduced by the presence of non-GR phase terms in the signal waveform when those are not present in the template. However, it is easy to appreciate that similar effects will be introduced by other possible mismatches (e.g. unknown large Post Newtonian orders, tidal effects, spins, eccentricity). It is therefore imperative for the GW community to concentrate on the development of as accurate waveforms as possible or of methods to be robust against the potential systematics that our approximate waveforms might introduce.
 
The uniqueness of the non-GR stealth bias, however, is that it cannot be eliminated with more precise numerical simulations or analytic models, as it represents the very uncertainty on our understanding of gravity in its strong-field regime. 

Furthermore, the conclusions drawn from studies that rely on the correct identification of the component objects of a compact binary source (\emph{e.g.} the measurement of the differential rate of coalescence in each class of systems or the measurement of the mass function) will have to be conditional on the assumption that GR is the correct description of the physics of the system. However, it is not farfetched to assume that if a deviation from GR is eventually detected, all inferences will be corrected \emph{a posteriori}.

In conclusion, the somewhat exotic, but with very real effects, ``stealth bias'' is nothing more than a consequence of our assumptions about the theory of gravity describing the process of gravitational radiation. This is a common phenomenon in every inference process, since the conclusions \emph{always} depend on the assumptions, but it is particularly worrying in the gravitational wave physics context, since it deals with the very foundations of our understanding of gravity.   

\section{Acknowledgments}

S.V. acknowledges the support of the National Science Foundation and the LIGO Laboratory. LIGO was constructed by the California Institute of Technology and Massachusetts Institute of Technology with funding from the National Science Foundation and operates under cooperative agreement PHY-0757058. 

W.D.P. is supported by the research program of the Foundation for Fundamental Research on Matter (FOM), which is partially supported by the Netherlands Organisation for Scientific Research (NWO). The work was funded in part by a Leverhulme Trust research project grant. 

The authors would like to acknowledge the LIGO Data Grid clusters, without which the simulations could not have been performed. Specifically, these include the computing resources supported by National Science Foundation awards PHY-0923409 and PHY-0600953 to UW-Milwaukee. Also, we thank the Albert Einstein Institute in Hannover, supported by the Max-Planck-Gesellschaft, for use of the Atlas high-performance computing cluster.
We would also like to thank Trevor Sidery, Christopher Berry, Neil Cornish, Andy Lundgren, Ilya Mandel, Chris Van Den Broeck, Ruslan Vaulin, Nico Yunes and Rai Weiss for useful comments and suggestions.
We are particularly grateful to Reed Essick for his comments and for carefully reading the manuscript.
This is LIGO document number P1300188.

\end{document}